\documentstyle[11pt,aaspp4]{article}
\slugcomment{Submitted to {\em The Astrophysical Journal}}

\lefthead{Southwell et al.}
\righthead{The supersoft X-ray source RX~J0513--69}

\newcommand{\etal}{\mbox{et\ al.\ }} 
\newcommand{\eg}{\mbox{e.g.\ }} 
\newcommand{\ie}{\mbox{i.e.\ }} 
\newcommand{\ergsec}{\,\mbox{$\mbox{erg}\,\mbox{s}^{-1}$}} 

\newcommand{\kmsec}{\,\mbox{$\mbox{km}\,\mbox{s}^{-1}$}}
 
\def\spose#1{\hbox to 0pt{#1\hss}} 
\def\simlt{\mathrel{\spose{\lower 3pt\hbox{$\mathchar"218$}}
     \raise 2.0pt\hbox{$\mathchar"13C$}}}
\def\simgt{\mathrel{\spose{\lower 3pt\hbox{$\mathchar"218$}}
     \raise 2.0pt\hbox{$\mathchar"13E$}}}

\begin{document}

\title{The nature of the supersoft  \\
     X-ray source RX~J0513--69}

\author{K.\,A.\ Southwell\altaffilmark{1}, M.\ Livio\altaffilmark{2}, 
P.\,A.\ Charles\altaffilmark{1}, D.\,O'Donoghue\altaffilmark{3} 
and W.\,J.\ Sutherland\altaffilmark{1}}

% Notice that each of these authors has alternate affiliations, which
% are identified by the \altaffilmark after each name.  The actual alternate
% affiliation information is typeset in footnotes at the bottom of the
% first page, and the text itself is specified in \altaffiltext commands.
% There is a separate \altaffiltext for each alternate affiliation
% indicated above.

\altaffiltext{1}{Dept.\ of Astrophysics, Nuclear Physics Building, 
Keble Road, Oxford, OX1~3RH, UK.\\ E-mail: kas@astro.ox.ac.uk; 
pac@astro.ox.ac.uk; wjs@astro.ox.ac.uk} 
\altaffiltext{2}{Space Telescope Science Institute, 3700 San Martin Drive, 
Baltimore, MD 21218, USA.\\ E-mail: mlivio@stsci.edu} 
\altaffiltext{3}{Department of Astronomy, University of Cape Town, 
Rondebosch 7700, South Africa.\\ Email: dod@uctvms.uct.ac.za}

% The abstract environment prints out the receipt and acceptance dates
% if they are relevant for the journal style.  For the aasms style, they
% will print out as horizontal rules for the editorial staff to type
% on, so long as the author does not include \received and \accepted
% commands.  This should not be done, since \received and \accepted dates
% are not known to the author.

\begin{abstract}

We present spectroscopy and photometry of the LMC supersoft binary system 
RX~J0513.9-6951. We derive a refined spectroscopic period of 
P$=0.761\pm0.004$~d, which is consistent with the value obtained from 
long term photometric monitoring (P$=0.76278\pm0.00005$~d). 
We see bipolar outflow components of He{\sc ii} and H$\beta$, with velocities 
of $\sim 3800$\,km\,s$^{-1}$, strongly suggesting that the compact object is 
a white dwarf. 
Using all the available optical and X-ray data, 
we construct a theoretical model to explain the principal features of the 
unusual variability of this source. In particular, we note that X-ray outbursts 
have only been seen at times of optical minima. From this, we conclude that 
the most likely cause of the X-ray outbursts is 
a photospheric contraction during a nuclear shell burning phase, rather than a 
thermonuclear flash or shocked emission. The system probably comprises  
a relatively massive white dwarf accreting at a high rate ($\sim 10^{-6} 
M_{\odot}~{\rm yr}^{-1}$) from an evolved donor star, and is observed close to 
pole-on. 

\end{abstract}

% The different journals have different requirements for keywords.  The
% keywords.apj file, found on aas.org in the pubs/aastex-misc directory, 
% contains a list of keywords used with the ApJ and Letters.  These are 
% usually assigned by the editor, but authors may include them in their 
% manuscripts if they wish. 


% That's it for the front matter.  On to the main body of the paper.
% We'll only put in tutorial remarks at the beginning of each section
% so you can see entire sections together.

% In the first two sections, you should notice the use of the LaTeX \cite
% command to identify citations.  The citations are tied to the
% reference list via symbolic KEYs.  We have chosen the first three
% characters of the first author's name plus the last two numeral of the
% year of publication.  The corresponding reference has a \bibitem
% command in the reference list below.
%
% Please see the AASTeX manual for a more complete discussion on how to make
% \cite-\bibitem work for you.   

\section{INTRODUCTION}

The supersoft X-ray sources (SSS) are a class of luminous
($L_{\mbox{bol}} \sim~10^{37} - 10^{38}$\ergsec) objects, with a
characteristic radiation temperature of $(1-10)~\times~10^5$\,K.
Seven such sources are now known in the Galaxy, 11 in the
Magellanic Clouds, 15 in M31 and candidates exist also in M101,
NGC~253 and M33 (see reviews by Hasinger 1994; Kahabka \&
Tr\"{u}mper 1996; Cowley \etal 1996).

The most popular theoretical model for the SSS consists of a binary
system in which a white dwarf accretes mass from a subgiant
companion at such a high rate that it burns hydrogen 
steadily at its surface (van den Heuvel \etal 1992; Rappaport
\etal 1994). The difference
between SSS and ordinary cataclysmic variables is in the high accretion rates 
(by a factor $\sim 100-1000$) in the SSS. 
More recently, Yungelson \etal (1995) have shown
that three major binary subpopulations may contribute to the
Galactic population of SSS with: (i)~low mass main sequence
donors, (ii)~low mass subgiant donors, and
(iii)~(super)giant donors. 

The transient supersoft source RX~J0513.9--6951 (hereafter 0513--69) 
was discovered in outburst during the ROSAT All Sky Survey
(Schaeidt, Hasinger \& Tr\"{u}mper 1993).  In 1990 
October/November the source brightened in X-rays by about a factor of 20
during a period of $\sim~10$ days.  A formal black body fit to
the PSPC data gave a temperature of about 40\,eV and a bolometric
luminosity of $\sim 2\times10^{38}$\ergsec, for a column density of 
N$_{\rm {\scriptsize H}} = 9.4 \times 10^{20}$~cm\,$^{-2}$. 
The source has
been monitored roughly every three months between November 1992
and October 1993 and was found to be three times in an off-state
and once, on 20/22 July 1993, in an on-state, with a similar brightness
to that observed in the first outburst (Schaeidt 1996). Furthermore, 
a series of weekly HRI pointings between 3~November 1994 and 3~March 1995 
revealed a third X-ray outburst around the end of December 1994. A typical 
turn-on time of $\sim 6$~d (from survey data) and turn-off time of 
$\simlt 1$ week (from the HRI pointings) 
were thus proposed for this source by Schaeidt (1996). 

This object has been identified with a $\sim 17$~mag
emission-line star in the LMC (Pakull \etal 1993; Cowley \etal 1993).
Spectroscopic observations by Crampton \etal (1996), 
hereafter C96, showed that the spectrum is dominated by
He~{\sc ii} emission lines and H$+$He{\sc ii} blends. Their 
radial velocity measurements suggested a binary period of 0.76~d, but no orbital
photometric variations were found. However, we have recently published a 
$\sim3$~year light curve of 0513--69, from observations made during the  
MACHO project (Alcock \etal 1996). This photometry revealed a high level of 
optical variability, including recurrent low states, in which the brightness 
drops by $\sim 1$~mag every $\sim 100-200$~days. The extended time-base over 
which these observations were made enabled us to detect an orbital 
modulation, although of small semi-amplitude ($\sim 0.02$~mag). The photometric 
period thus derived, P$=0.76278\pm0.00005$~d, is consistent with the 
independent spectroscopic result.  

In the current work, we present further spectroscopic and photometric
observations of 0513--69, in which we see some unusual features, including 
evidence for high velocity outflows and optical variability on timescales as 
short as $\sim 3$~h. 
We then 
use all the available optical and X-ray data in an attempt to
construct a comprehensive theoretical model for the source. The 
spectroscopic observations are described in \S~2, and the photometry 
in \S~3. In \S~4, we present a refined spectroscopic period, and 
consider the implications of the derived binary parameters and mass function. 
A discussion and the proposed model follow.

\section{SPECTROSCOPY}

\subsection{Observations and Reduction}

We obtained intermediate resolution spectra of 0513--69 on the nights 
29/11/94 -- 1/12/94 using the 3.9\,m Anglo-Australian Telescope at Siding 
Spring, Australia. The detector was a Tek CCD attached to the RGO 
spectrograph, using 
a 1200V grating. The wavelength coverage is 4367--5070\AA\ in the blue and 
6103--6857\AA\ in the red, with a resolution of $\sim1.3$\,\AA. 
Due to variable cloud and generally poor seeing, no standard 
stars were observed, hence the spectra are not flux calibrated. 
Cu-Ar arc spectra were taken at regular intervals for 
calibration of the wavelength scale in the reduction procedure. 
A series of tungsten flat fields and bias frames were also obtained.
Tab.~1 lists the journal of observations. 

The data reduction was performed using the Starlink {\sc figaro}
package and {\sc pamela} routines of K.\,Horne.  Removal of the bias
signal was achieved through subtraction of the mean overscan level on
each frame. This was acceptable since an examination of the bias
frames showed no significant structure. Small scale pixel-to-pixel
sensitivity variations were removed by multiplying
by a balance frame prepared from flat fields of a tungsten lamp.
One-dimensional spectra were extracted using the optimal algorithm of
Horne (1986), and calibration of the wavelength scale was achieved
using the {\sc molly} package of T.\,R.\ Marsh.

% Authors may indicate to the editorial staff where they would like 
% figures and tables to be placed in the manuscript.  This is done with
% either the \placefigure{KEY} or \placetable{KEY} commands.  These
% commands require \label{KEY} commands to be placed appropriately with
% corresponding table and figure captions.  When the manuscript is
% printed a short note is printed on the page where the figure or table
% is to go.  These commands are ignored in the aaspp4 and aas2pp4 styles.

%\placetable{tbl-3}
%\placefigure{fig1}

% In this section, we see the use of the \subsection command to set off
% an independent subsection.  We only have one here; usually there would
% be several.

\subsection{Results and Analysis}

We present below the spectroscopic data and give a 
description of the principal features. A detailed radial velocity analysis 
is given later in \S~4. 

We show in Fig.~1 the variance-weighted average of all our blue spectra. 
He{\sc ii} and H$\beta$ are seen strongly in emission, with narrow cores and 
broad bases. Previous spectroscopic observations (Pakull \etal 1993; 
C96) have shown the presence of highly ionised elements, 
notably O{\sc v} (5595\,\AA), 
O{\sc vi} (5290\,\AA) and N{\sc v}. Given our limited spectral coverage, 
we are unable to comment on the presence of oxygen. However, we find marginal 
evidence for the 
N{\sc v} 4603/4619~\AA\ emission lines reported by the latter authors. The 
spectral region around $4600-4700$\,\AA, shows evidence for the 
Bowen N{\sc iii}-C{\sc iii} $\lambda\lambda~4640-50$~\AA\ complex and probable 
C{\sc iv} 4658.3\AA, although these lines are contaminated by the underlying 
He{\sc ii} emission. 

In Fig.~2, we present the mean red spectrum, which is dominated by 
strong H$\alpha$ emission. Although the signal-to-noise ratio is 
poor, we are able to identify components (marked S$^-$ and S$^+$, following 
the nomenclature of C96) in 
emission. Similar features are evident on each side of the He{\sc ii} 4686\AA\ 
and H$\beta$ lines in Fig.~1. 
These unusual features were 
first noted in the blue 
by Pakull (1994) and Cowley \etal (1996), and 
interpreted as high velocity components of He{\sc ii} and H$\beta$. 
Recently, C96 have confirmed this identification, being 
unable to attribute the lines to any likely ionic species, and 
concluding an origin in some type of bipolar outflow. 

All the spectra were examined for nightly variations in the strength and shape 
of the emission lines. However, 
no significant changes were apparent in the major lines 
over the course of 
the observing run. In Tab.~2, we list the mean 
equivalent widths of the major emission features for each night, and also for 
the overall average red and blue spectra. We give the 1$\sigma$ statistical 
errors, although we caution that 
systematic uncertainties may arise due to the difficulty of assessing the 
continuum 
level for broad based line profiles. Also tabulated are the equivalent widths 
of the S$^+$/S$^-$ emission components. 
The wavelengths and relative velocities of the S$^+$/S$^-$ 
components, measured from the average spectra of Figs.~1 and 2, 
are listed in Tab.~3. 
Our findings are consistent 
with the measurements of C96, supporting the interpretation 
of these features as Doppler-shifted components of the He{\sc ii} and 
Balmer lines. 

We see further evidence for high velocity flows within this system in 
the emission line profiles of He{\sc ii} 4686\AA\ and 
H$\beta$ (see Fig.~1). The former line exhibits 
a pronounced blue wing, extending to $\sim 3800$~km\,s$^{-1}$ (see also 
C96). Furthermore, the blue side of the H$\beta$ emission appears to be 
absorbed.

\section{PHOTOMETRY}

\subsection{The Optical Light Curve}

0513--69 lies in one of the fields surveyed by the MACHO 
project (see \eg Alcock \etal 1995). As a result of this serendipitous 
photometric monitoring, a record of the optical variability of this source 
exists from 1992 August to the present time. 
For completeness, 
we show in Fig.~3 the MACHO light curve of 0513--69 for the period 
1992 Aug.~22 -- 1995 Nov.~27. 
These observations have recently been 
published in Alcock \etal (1996), to which we refer the reader for further 
details regarding the data acquisition and reduction, and a discussion 
of the variability. However, we have also marked on Fig.~3 arrows indicating 
the times at which the system was known to be definitely on and 
off in X-rays; we refer to this later in \S~5.3 and 
\S~5.4. 
The optical light curve is dominated by recurrent low states, representing 
$\sim$~0.8~mag reductions in intensity, which 
occur at quasi-regular 
intervals. In addition, 0513--69 exhibits erratic variability on shorter 
timescales, with fluctuations of up to $\sim 0.3$~mags occurring on timescales 
of days. 

We have also obtained white-light photometry of 0513--69 
using the SAAO 1~m telescope at Sutherland, South Africa. These 
observations were made on 1995 Feb.~5 ($\equiv$~JD 2449754 - marked ``SAAO'' on 
Fig.~3), at a time when 0513--69 was in an optical high state. 
The detector was the UCT 
CCD, operating in frame transfer mode. In this configuration, only half of 
the chip is exposed, and at the end of the integration, the signal is read 
out through the masked half. In this way, we are able to obtain consecutive 
integrations with virtually no CCD dead time. The effective wavelength of 
the instrument, in the absence of a filter, is approximately that of the 
Johnson V passband, but with a bandwidth of $\sim 4000$\AA\ (FWHM). 
In Fig.~4, we show the resulting 
light curve, which consists of 1328 10~s exposures, obtained 
over a continuous 220~min period. The magnitude of 0513--69 and a comparison 
star are plotted relative to a local standard. We see a fading in the 
brightness of 0513--69 of $\sim 0.1$~mags, which is not observed in the 
comparison star. Thus, we find evidence that 
0513--69 can exhibit significant optical variability 
on timescales as short as $\sim 3$~h. We performed a power spectrum analysis of 
the data, using the method of Scargle (1982). No significant power was revealed 
within the range of periods to which the search was sensitive 
($\sim$ 30\,s~$-$~2\,hr). 
By adding an artificial sinusoidal component to the data, we derive an 
upper limit of $\sim 0.015$~mags for the semi-amplitude of any real modulation 
on these short timescales.

\section{BINARY PARAMETERS}

\subsection{The Spectroscopic Period}

\subsubsection{Measurement of line velocities}

We measured the line velocities of He{\sc ii}~4686 by fitting 
triple Gaussian functions to the line profiles. 
These fits consisted of a narrow component for the line core, an 
underlying broader Gaussian to account for the red emission shoulder 
and a further wide, low-amplitude component, to fit 
the extended blue emission wing, which includes the 
$\lambda\lambda~4640-50$~\AA\ Bowen blend. We list in Tab.~4 the radial 
velocities of He{\sc ii}~4686 for our 10 blue spectra, obtained 
from the strong narrow component. The velocity centroids of the broad 
components were far less accurately determined, and we do not give these. 
A triple Gaussian fit to the He{\sc ii}~4686 profile of the average blue 
spectrum is shown in Fig.~5. 

\subsubsection{Period search}

The periodogram for He{\sc ii}~4686 is shown in 
Fig.~6. The frequency space ranges from $0.1-10.0$~cycles per day, with a 
search interval of 0.001~cycles per day 
(a more wide ranging search failed to reveal further significant power). 
We have 
applied the {\sc clean} algorithm of Roberts, Leh\'{a}r and Dreher (1987),  
which essentially deconvolves the spectral window from the data. 
The major peak is at 
P$=0.761\pm0.025$~d, which is entirely 
consistent with the ``best'' spectroscopic 
value of 0.76~d obtained by C96. The power arising at 4.54~d 
($\nu = 0.22$~cycles~d$^{-1}$) is considered due to a one day$^{-1}$ 
alias. 

Given the extremely limited sampling of our dataset, we combined our 
He{\sc ii}~4686 radial velocities with those of C96, which were obtained only 
$\simlt 3$~weeks before our observations.  
In order to correct for 
any systematic differences between the data, we first subtracted the 
mean value from each set. The CLEANED periodogram, shown in Fig.~7,  
is dominated by a peak 
at P$=0.761\pm0.004$~d, with one day$^{-1}$ aliases at 
$\nu \approx 0.33$~cycles~d$^{-1}$ and $\nu \approx 2.30$~cycles~d$^{-1}$.  

We therefore confirm the spectroscopic period suggested by C96, 
and find that 
this is consistent, within the 1$\sigma$ errors, with the photometric period 
of P$=0.76278\pm0.00005$~d (Alcock \etal 1996). 
 
\subsection{The Mass Function and Parameter Space}

We folded our He{\sc ii}~4686 radial velocities 
on P$=0.76278$~d (this being the most 
accurate period determination available) using the photometric 
ephemeris for the time of maximum optical brightness: 
T$_{\rm o}={\rm JD}\,2448857.832(5) + 0.76278(5)E$ (Alcock \etal 1996). 
The resulting radial velocity curve is shown in Fig.~8 with a sinusoidal fit. 
We obtain a velocity semi-amplitude, K$=14.5\pm3$\,km\,s$^{-1}$, 
a systemic velocity, $\gamma=297\pm1$\,km\,s$^{-1}$ and zero crossing 
phase $\phi_{\rm o}=0.17\pm0.09$. The error on the latter quantity includes 
the uncertainty in the photometric ephemeris. We also combined our velocity 
data with that of C96 (see \S~4.1.2 
above) and repeated this procedure, although the data were difficult to fit 
since C96 give no errors. 
This yielded a zero-crossing phase of 
$0.30\pm0.08$ and velocity semi-amplitude of $\sim 11$\,km\,s$^{-1}$, which 
are not significantly different from the parameters obtained from our data 
alone. Given 
the very low amplitude of the modulation and the intrinsic uncertainty in 
the photometric ephemeris, we are unable to obtain an accurate value for 
$\phi_{\rm o}$. 
Thus, it is difficult to draw conclusions from the relative phasing of 
the light and velocity curves. However, the data do suggest that at maximum 
optical light, the velocities are most blue-shifted. Assuming that the 
He{\sc ii} arises near the compact object, this could be explained if we are 
viewing the inner irradiated edge of the disk, near the impact point of the 
stream. 

The mass function for the parameters derived from our data alone is 
f\,(M) = PK$^3$/2$\pi$G = 0.0002$\pm0.0001$ \,M$_{\odot}$, which is 
plotted in Fig.~9 for various inclinations (see also C96). Clearly, if the 
accretor is a white dwarf, the inclination of the system must be very low, 
\ie seen close to pole-on. 
This conclusion is unaltered for the highest mass function allowed by the 
uncertainty on the measurement. 
Whilst it appears statistically unlikely that we are 
viewing the system close to pole-on ($i \simlt 7^{\circ}$), the following 
points argue for this possibility (see also C96):

(1)~If the inclination is not very small (see above), the brightness of the
disk would imply unreasonably high accretion rates (in excess of the 
Eddington limit - see \S~5.2).

(2)~The orbital photometric variations are extremely small ($\sim 0.04$~mag 
full amplitude - see \S~1).  With a high accretion rate (\S~5.2) and 
steady burning 
at the white dwarf surface (\S~5.4), the illuminated hemisphere of the
secondary would be expected to produce larger variations at higher
inclination angles. For example, the eclipsing SSS, CAL~87, exhibits 
photometric variations of $\sim 0.5$~mag (excluding the eclipse) 
for an inclination of $i \simgt 70^{\circ}$ 
(Schmidtke \etal 1993; Schandl, Meyer-Hofmeister \& Meyer 1996). 

We note that in such a low inclination system, in which the accretor is 
a white dwarf, 
the implied mass of the companion star is such that it would not fill its 
Roche lobe if on the main sequence. An evolved donor star is therefore 
required. 

\section{DISCUSSION AND MODEL}

We will now discuss the implications of the observations
presented here, as well as of other available data, for models of
the system.

\subsection{The Nature of the Compact Object}

The first thing we would like to establish is the nature of the
compact object. To this goal, we note that observations of young
stellar objects (\eg Reipurth \& Heathcote 1993), of AGN (\eg
Blandford 1993), of SS~433 (\eg Vermeulen 1993) and of the Galactic black hole 
candidates GRS~1915+105 (Mirabel \& Rodriguez 1994) and GRO~1655-40 
(Hjellming \& Rupen 1995) {\it all}
indicate that the velocities of jets are always of the order
of {\it the escape velocity from the
central object}.  The observations presented in \S~2 (see also
C96) indicate bipolar outflows or jet speeds of
$V_{\rm bipolar} \sim~3800$\kmsec. If interpreted as an escape
velocity, the observed value of $V_{\rm bipolar}$ corresponds to a value
of the mass to radius ratio of the compact object of $M/R~\sim
40~M_{\odot}/R_{\odot}$, {\it which is typical for a white dwarf}.
Indeed, outflows with velocities of this order have been observed in 
cataclysmic variables (\eg Drew 1991; Drew, Hoare \& Woods 1991). 

Since the value of $M/R$ identifies the accreting object unambiguously as a 
white dwarf, we check further if the observed luminosity is consistent with a 
white dwarf burning hydrogen in a shell. For such an object, the luminosity 
is related to the white dwarf mass by (\eg Iben \& Tutukov 1989)
\begin{equation}
L/L_{\odot} \simeq 4.6\times10^4 (M_{\rm WD}/M_{\odot} - 0.26).
\end{equation}
Note that this relation differs somewhat from the usual
Paczy\'{n}ski-Uus relation (Paczy\'{n}ski 1970; Uus 1970) which is
appropriate for AGB stars which have both hydrogen and helium
burning shells. The bolometric luminosity of 0513--69 is probably between 
$9.5\times10^{37}$\ergsec (obtained from an 
LTE model atmosphere fit; Reinsch \etal 1996) and 
$2\times10^{38}$\ergsec (obtained assuming a black body fit; Schaeidt,  
Hasinger \& Tr\"umper 1993). 
Using Eqn.~1 with these values gives a mass of $\sim 0.8-1.4\,M_{\odot}$, again 
consistent with a white dwarf. 

From the above discussion we therefore conclude that the compact
object in 0513--69 is almost certainly a white dwarf.

\subsection{The Accretion Rate}

Much of the optical luminosity of the system is probably
generated in the accretion disk (although some fraction may
represent the effects of nuclear burning and of illumination
of the accretion disk and
the secondary star by the hot white dwarf).  For a distance modulus of
18.5 to the LMC and $E_{B-V} = 0.1$ (\eg Panagia \etal 1991), we
obtain (in the optical high states) $M_V~\simeq~-2$. {\it If} we assume that 
the entire luminosity comes from a standard accretion disk then, using the
fact that the luminosity of the accretion disk is given
approximately by (Webbink \etal 1987)
\begin{equation}
M^{\rm disk}_V \simeq -9.48 - \frac{5}{3} \log
\left( \frac{M_{\rm WD}}{M_{\odot}} \frac{\dot{M}}{M_{\odot} {\rm yr}^{-1}}
\right) - \frac{5}{2} \log~(2\cos i)~,
\end{equation}
where $\dot{M}$ is the accretion rate and $i$ is the inclination angle,
we obtain for $M_{\rm WD} = 1M_{\odot}$ and $i \simeq 10^{\rm o}$
(see \S~4.2), an accretion rate of $\dot{M}~\simeq~10^{-5}
M_{\odot}~{\rm yr}^{-1}$. This accretion rate is of the
order of the Eddington
value, and it indicates that at least some fraction of
the optical light is due to illumination (and perhaps nuclear 
burning on the white dwarf surface).
It probably remains true, however, that the  
accretion rate in this system is extremely high. 
Another indication of the fact that the accretion rate in 0513--69 may be 
higher than in other
similar systems (\eg CAL~83) comes from the observation of the
bipolar outflow (\S~2.2 and \S~5.1). Typically, the mass flux in
jets is of the order of 1--30\% of the disk accretion rate (\eg
Lizano \etal 1988), and while not all the objects which exhibit
jets have luminosities near the Eddington limit (although SS~433
does), it is certainly the case that the intermittent nature of the
jets in young stellar objects seems to be associated with
episodes of an increased accretion rate through the disk (\eg
Reipurth \& Heathcote 1993).

The observed drops in the optical
luminosity, by $\sim 0.8$~mag (\S~3), if interpreted as a
reduction in the accretion rate, correspond to a decrease in
$\dot{M}$ by a factor $\sim$~3. We should note, however, that if
the optical luminosity is actually dominated by reprocessed radiation from
the accretion disk and/or the secondary star, both being irradiated by the
steady burning white dwarf, then the decrease in $\dot{M}$ could be by a 
larger factor.
Such occasional drops in the accretion rate 
are observed in many nova-like variables and in particular in the group of 
cataclysmic variables known as VY
Scl stars (\eg Shafter 1992; Robinson \etal 1981; Honeycutt 1995).
Similarly to 
the case of 0513--69, the downward transitions in VY Scl stars
also occur on a timescale of tens of days (\eg Hudec, Huth \&
Fuhrmann 1984; Liller 1980; Rosino, Romano \& Marziani 1993;
Honeycutt 1995).

Further evidence for the relation between the bipolar outflows
and the accretion rate may be inferred from comparison of spectroscopic 
observations obtained during both high and low states. 
Optical spectroscopy obtained during a low state in December 1993 (C96; 
Reinsch \etal 1996) revealed the equivalent widths of the emission lines to 
be significantly weaker than in high state observations. 

We shall return to the question of what can cause the drops in
the mass transfer rate when we discuss a comprehensive model for
the system in \S~5.4 below.

\subsection{The Cause of the X-ray Outbursts}

There are four main ways in which an accreting white dwarf can exhibit
transient X-ray outbursts: (i)~the outbursts may represent
some phases in thermonuclear flashes occurring when a
critical mass for ignition
is accumulated (\eg Iben 1982; Prialnik \& Kovetz 1995), examples 
of this behaviour being provided by classical novae; 
(ii)~the
X-ray outbursts may be the consequence of a photospheric
contraction during a nuclear shell burning phase (even if the
burning was steady; \eg MacDonald,
Fujimoto \& Truran 1985; \"Ogelman \etal 1993; Krautter \etal 1996); 
again, this behaviour
has been observed in some nova systems;
(iii)~the X-ray luminosity may be generated in shocks
resulting from the interaction of an ejected shell with the ISM; (iv)~the 
outbursts may simply represent epochs in which the X-ray source is not 
shielded by intervening material (\eg Pakull 1996). 

Examining the X-ray data {\it alone} it is difficult to rule out
regular thermonuclear flashes as a possible cause for the
recurrent X-ray outbursts (this was, in fact, the model proposed by
Kahabka 1995). 
However, when the X-ray data are examined {\it together} with the
optical data, flashes caused by the accumulation of a critical
mass become extremely unlikely.  The reason is simply
the fact
that the second and the third 
X-ray outbursts observed in the system in July
1993 and in December 1994
occurred {\it during optical minima} (see Fig.~3), while
{\it no outbursts were observed during optical high states} (Schaeidt 1996; 
Pakull \etal 1993).  If this represents the rule (rather than
being an accident), then it is very difficult to reconcile
with a thermonuclear flash model, which is normally accompanied by
radius expansion and an increased optical luminosity (see \eg Livio 1994 for a 
discussion). Even if a thermonuclear flash was able to produce a decrease in 
the optical 
luminosity (for example, if it led only to a modest expansion, but which 
nevertheless managed to destroy the disk), this would {\it follow} the X-ray 
rise, contrary to what is probably observed (see \S~5.4 and Fig.~3). 
We may thus rule out the thermonuclear flash model. 
The shock emission model ((iii) above) may also be eliminated as a 
possible source for the increase in the X-rays, since the ejection of a shell
is normally a consequence of significant expansion (\eg Prialnik \& Kovetz 
1995). 

We are therefore left with 
the episodic unveiling of a permanent X-ray source or 
contraction during a steady shell burning phase 
as the most probable origins for the increase in the X-ray luminosity (see 
Pakull \etal 1993; Pakull 1996). 
In the former model, the white dwarf is burning steadily 
all the time, but is visible (in X-rays) only during phases of 
low mass transfer rate (see \S~5.4 for a possible cause of such phases). At 
other times, the wind and bipolar outflow are 
optically thick to soft X-rays. It should be noted that the 
decrease in the optical should precede the rise in the X-rays, which is 
probably consistent with the available data (see \S~5.4). 

We may 
investigate whether the outflow can indeed shield the source using the 
following simplified approximation. If we assume a spherical outflow, 
characterised by a velocity law of the type $V=V_{\infty} \left ( 
1 - R_{\rm WD}/r \right )^{\beta}$, where $V_{\infty}$ is the outflow velocity 
at a large distance, then we can calculate the optical depth in the outflow, 
\begin{equation}
\tau = \frac{\sigma}{4\pi m_p V_{\infty} R_{\rm WD}} f(X, \beta) \dot{M}_{\rm 
outflow}. 
\end{equation}
Here, $m_p$ is the proton mass, $\sigma$ is the absorption cross-section to the 
soft X-rays, $\dot{M}_{\rm outflow}$ is the mass outflow rate and $f(X, \beta)$
is a function of $X \equiv R_{\rm WD}/R_{\rm S}$, where $R_{\rm S}$ is the 
sonic radius in the outflow and $\beta$ is the exponent in the velocity law. 
For values of $\beta$ in the range $1.0-4.5$ and $X \sim 1/50$, both typical 
for nova-like variables (\eg Knigge 1995), $f(X, \beta) \sim 0.02$. If we 
use, for example, the photoelectric absorption cross-section at $0.28-0.40$~keV 
from Morrison \& Mc\,Cammon (1983), $V_{\infty} \sim 4000$~km\,s$^{-1}$, and 
$R_{\rm WD} \sim 10^9$~cm, we obtain that the source will be totally obscured 
($\tau \sim 5$) for $\dot{M}_{\rm outflow} \simgt 3 \times 10^{-9} 
M_{\odot}$~yr$^{-1}$. This number should not be taken as representing the real 
value, since the actual obscuration depends crucially on the geometry of the 
real outflow from the accretion disk (\eg Knigge, Woods \& Drew 1995), 
especially in a nearly pole-on system. 

Whilst the above discussion does indicate that a scenario 
of this type is, in principle, viable, we find the white dwarf photospheric 
contraction model more reasonable; this is discussed fully in 
\S~5.4 below. It is important to note that the appearance 
of a relatively short lived X-ray phase due to contraction of the
photosphere, during shell burning, has been established
observationally for both GQ Mus (\"{O}gelman \etal 1993; Shanley
\etal 1995) and V1974~Cyg (Krautter \etal 1996). This puts us now in
a position where we can attempt to propose a comprehensive model
for the system, taking all the available observational data into
consideration.

\subsection{A Comprehensive Model of Photospheric Contraction}

The principal features of the photospheric contraction model which 
emerge as a result of the discussion in the
previous sections are the following. The system consists of a white dwarf
which may be fairly massive (both because of the constraints
imposed by the accretion luminosity, see \S~5.2, and by the
inclination angle, \S~4.2; see also below), which accretes from an 
{\it evolved} companion. 

The accretion rate is normally very high (perhaps $\sim10^{-6}
M_{\odot}~$yr$^{-1}$), at a value which is near the top of the
steady burning strip in the $\dot{M} - M_{WD}$ plane
(\eg Nomoto 1982).  Under these conditions,
the white dwarf is slightly inflated (by perhaps no more than a factor 
$\sim3$ in radius; Kovetz \& Prialnik 1994), and most of the shell luminosity
is probably emitted in the UV.  The mass transfer rate, and
concomitantly the optical luminosity, suffer occasional drops by
about a factor~3 or more.  This is a very similar phenomenon to the one
exhibited by VY Scl stars and some nova-like variables (\eg Honeycutt
1995; Honeycutt, Robertson \& Turner 1995).

The important point here is that 0513--69, like the VY Scl
stars, experiences only {\it downward} transitions.  Livio \& Pringle
(1994) suggested a model for VY Scl
stars, in which the reduced mass transfer rate is a consequence
of a magnetic spot covering the $L_{1}$ region.  VY Scl stars are
normally found in the period range 3--4 hrs and in the Livio \&
Pringle model, this is a consequence of the fact that when the
rotation rate of the star (which is coupled to the orbit)
increases, so does the magnetic activity and the fraction of the
stellar surface covered with spots (\eg K\"urster \etal 1992).
One may therefore wonder why the secondary in 0513--69, which has
an orbital period of $\sim~0.76$~d, should exhibit
a similar behaviour. However, it should be remembered that the
secondary in 0513--69 is evolved. The physical quantity which
characterizes the magnetic activity is the Rossby number, $P_{\rm
rot}/\tau_{\rm C}$, where $\tau_{\rm C}$ is the convective overturn
time in the envelope (\eg Schrijver 1994). 
Since evolved stars have deeper convective
envelopes (longer $\tau_{\rm C}$) than main sequence stars, they
exhibit the same level of activity at longer $P_{\rm rot}$ (see
\eg Schrijver 1994 for a review). In fact, the
orbital period of 0513-69 is close to the range spanned by the magnetically 
active RS~CVn stars. 

Once the accretion rate drops, the photosphere contracts slightly
(\eg Kovetz \& Prialnik 1994; Kato 1985), raising the effective
temperature and thus producing an increase in the X-ray
luminosity. It is 
important to note in this respect that the decrease in the optical luminosity, 
{\em precedes} the rise in the X-rays. Fig.~3 shows the times at which 
the X-rays were known to be off (arrows marked ``NX'') and on 
(arrows marked ``X''), in 
relation to the optical behaviour. Using the fact that the X-ray rise time 
is of order 6~days (Schaeidt 1996 - see also \S~1), we may be fairly confident 
that, at least in the X-ray outburst of December 1994 (day number $\sim$\, 
1715 in Fig.~3), the optical had already started to fall before the X-rays 
turned on. In this respect, the episodic unveiling model (discussed in \S~5.3 
above) is also consistent with the available data. However, Reinsch \etal 
(1996) 
have shown that the optical low states of 0513--69 are accompanied by a 
reddening of $\sim 0.1$~mags, which is quantitatively consistent with the 
decreased disk illumination resulting from a contraction in the white dwarf 
radius. 

Furthermore, the photospheric contraction model is able to 
provide a natural explanation of details in the light curve morphology, as 
described below. 
The increased X-ray flux irradiates
the companion star and, either by inflating material above the secondary's 
photosphere and causing it to be transferred
(\eg Ritter 1988), or by heating the magnetic spot area (which
is generally cooler, \eg Parker 1979), causes the mass transfer
rate to increase again.  This produces the step-like behaviour of
the optical light curve when the luminosity increases, or the
small jump in the optical luminosity in the middle of the low
state (around day 1539, Fig.~3). The following should also be noted: in 
{\it all} cases, the luminosity immediately following the end of 
one of these low states 
is slightly higher than the average between these events (see Fig.~3). This is 
entirely consistent with the above model, since the temporary blocking of 
the mass transfer (\eg by a magnetic spot), plus the effect of the 
irradiation of the secondary by the X-rays, is likely to result in a somewhat 
increased mass transfer rate, once the blocking is removed. 

A question which needs to be addressed is whether the contraction and 
expansion of the white dwarf photosphere can occur on the observed timescale 
($\sim$ 1 week). An examination of the results of Kovetz \& Prialnik (1994) and 
Kato (1996) reveals that this is indeed possible, if the white dwarf is 
massive ($M_{\rm WD} \sim 1.3-1.4~M_{\odot}$). This can be easily understood if
we realise that the contraction timescale can be reasonably approximated by 
the duration of the mass-ejection phase (Livio 1992): 
\begin{equation}
\tau_{\rm duration} \simeq \xi \left( \frac{M_{\rm WD}}{M_{\rm C}}
\right)^{-1} \left[ \left( \frac{M_{\rm WD}}{M_{\rm C}} \right) 
^{-2/3} - \left( \frac{M_{\rm WD}}{M_{\rm C}} \right) ^{2/3}
\right] ^{3/2},
\end{equation}
where $\xi$ is nearly a constant and $M_{\rm C}$ is the Chandrasekhar
mass. 
Using a value for 
$\xi \approx 51$~days, which fits the observations of 84 novae in M31 and 15 
novae in the LMC (Della Valle \& Livio 1995), gives a contraction timescale 
of less than 4 days for a $1.3 M_{\odot}$ white dwarf. 

Assuming that the main ingredients of the model outlined above
are correct, we may ask what is the cause for the difference in the
X-ray behaviour of 0513--69 and similar sources (such as CAL~83).  
The main difference is probably in the ratio of the 
mean mass transfer rate to the rate at which steady burning occurs (for the 
given white dwarf mass). This ratio is  
probably higher for 0513--69, a hypothesis which is supported by the greater 
optical luminosity of this system ($V_{\scriptsize mean} \sim 16.8$ in the 
high states, 
compared with $V_{\scriptsize mean} \sim 17.3$ for CAL~83; Cowley \etal 1993). 
Evidence for a high mass transfer rate is provided both by the
brightness of the accretion disk, and by the fact that a bipolar
outflow is not observed in CAL~83 (C96).

Finally, we should point out that the main model discussed in the present work 
(\S~5.4) is based on the probable 
observation that the rise in X-rays came {\it after} the optical luminosity 
was observed to drop. 
Thus, it 
can be tested directly by long-term monitoring of the system in the
optical and X-ray regimes. In particular, the model predicts that
increases in the X-ray luminosity should follow drops (by $\sim$~1~mag) in
the optical luminosity. 
If future observations will show that this is indisputably the case, then the 
model of photospheric contraction will be further strengthened.

\acknowledgements

We are grateful to the MACHO collaboration for allowing us access to 
their long term photometry of 0513--69. We thank the staff at the AAT, Siding 
Spring, for their assistance with the spectroscopic observations. 
KAS acknowledges the hospitality of ST~ScI, and support from PPARC. 
ML acknowledges 
support from NASA Grants NAGW~2678 and GO-4377. 
We are grateful to Paul Schmidtke and Anne Cowley for providing us with
information about their observations, and to the anonymous referees for their 
detailed comments.

\clearpage
 
\begin{deluxetable}{cccc}
\tablecaption{Journal of observations. \label{tbl-1}}
\tablewidth{0pt}
\tablehead{
\colhead{Date (m/d/y)} & \colhead{HJD$-$2440000} & \colhead{Exposure (s)} & 
\colhead{Wavelength coverage (\AA)} }
\startdata
11/29/94 & 9686.214   & 1000  & 4367$-$5070 \nl
11/29/94 & 9686.226  & 1000 & `` \nl
11/29/94 & 9686.237  & 600 & 6103$-$6857 \nl
11/29/94 &  9686.245 & 600 & `` \nl
11/29/94 & 9686.253 & 600 & `` \nl
11/30/94 & 9687.023  & 1800 & 4367$-$5070  \nl
11/30/94 & 9687.039 & 855 & `` \nl
12/1/94 & 9688.019  & 1800 & `` \nl
12/1/94 & 9688.070  & 1800 & `` \nl
12/1/94 & 9688.093  & 1800 & `` \nl
12/1/94 & 9688.115  & 1800 & `` \nl
12/1/94 & 9688.139  & 1800 & `` \nl
12/1/94 & 9688.161  & 1800 & `` \nl
\enddata
\end{deluxetable}

\begin{deluxetable}{crccc}
\tablecaption{Mean equivalent widths of emission features (\AA)  \label{tbl-2}}
\tablewidth{0pt}
\tablehead{
\colhead{Line} & \colhead{Night 1} & \colhead{Night 2} & 
\colhead{Night 3} & \colhead{Overall average} }
\startdata
He{\sc ii}~4541 & 1.2$\pm$0.3 & 1.5$\pm$0.2 & 1.1$\pm$0.1 & 1.4$\pm$0.1 \nl
He{\sc ii}~4686 & 20.7$\pm$0.8 & 19.2$\pm$0.5 & 20.6$\pm$0.2 & 20.4$\pm$0.2 \nl
S$^{-}_{{\scriptsize 4686}}$ & 0.9$\pm$0.3 & 0.5$\pm$0.2 & 0.9$\pm$0.1 & 0.8$\pm$0.1 \nl
S$^{+}_{{\scriptsize 4686}}$ & 0.3$\pm$0.3 & 1.0$\pm$0.2 & 1.1$\pm$0.1 & 1.1$\pm$0.1 \nl
H$\beta$ & 13.2$\pm$0.6 & 11.3$\pm$0.4 & 10.7$\pm$0.2 & 10.8$\pm$0.1 \nl
S$^{-}_{{\rm {\scriptsize H}}{\scriptsize \beta}}$ & 0.1$\pm$0.3 & 0.0$\pm$0.2 & 0.2$\pm$0.1 & 
0.2$\pm$0.1 \nl
S$^{+}_{{\rm {\scriptsize H}}{\scriptsize \beta}}$ & 1.3$\pm$0.4 & 0.8$\pm$0.3 & 1.4$\pm$0.1 & 
1.3$\pm$0.1 \nl
H$\alpha$ & 48.2$\pm$1.4 & -  &  - & 48.2$\pm$1.4 \nl
S$^{-}_{{\rm {\scriptsize H}}{\scriptsize \alpha}}$ & 1.6$\pm$0.6 & - & -  & 
1.6$\pm$0.6 \nl
S$^{+}_{{\rm {\scriptsize H}}{\scriptsize \alpha}}$ & 2.6$\pm$0.7 & -  & -  & 
2.6$\pm$0.7 \nl
\enddata
\end{deluxetable}

\begin{deluxetable}{cccccc}
\tablecaption{Approximate wavelengths and relative velocities of 
Doppler-shifted components.  \label{tbl-3}}
\tablewidth{0pt}
\tablehead{
\colhead{Line} & \colhead{$\lambda_o$ (\AA)} & \colhead{S$^{-}$ (\AA)} & 
\colhead{S$^{+}$ (\AA)} & \colhead{S$^{-}$ (km\,s$^{-1}$)} &
\colhead{S$^{+}$ (km\,s$^{-1}$)}  }
\startdata
He{\sc ii}~4686 & 4691 & 4631 & 4752 & 3850 & 3900 \nl
H$\beta$ & 4866 & 4806 & 4932 & 3700 & 4050   \nl
H$\alpha$ & 6569 & 6488 & 6657 & 3700 & 4000 \nl
\enddata
\end{deluxetable}

\begin{deluxetable}{ccc}
\tablecaption{Radial velocities of He{\sc ii}~4686. \label{tbl-4}}
\tablewidth{0pt}
\tablehead{
\colhead{Spectrum no.} & \colhead{HJD$-$2440000} & \colhead{Velocity 
(km\,s$^{-1}$)} }
\startdata 
1  & 9686.214 & 291$\pm$5  \nl
2 &  9686.226 & 272$\pm$7   \nl
3  & 9687.023 & 288$\pm$3 \nl
4  & 9687.039 & 293$\pm$10 \nl
5  & 9688.019 & 313$\pm$7 \nl
6  & 9688.070 & 312$\pm$3 \nl
7  & 9688.093 & 311$\pm$2 \nl
8  & 9688.115 & 312$\pm$5 \nl
9  & 9688.139 & 309$\pm$2 \nl
10  & 9688.161 & 307$\pm$3  \nl
\enddata
\end{deluxetable}

% Text for table footnotes follows the tabular data and must be inside the
% deluxetable environment.  Note that it is OK to put \ref's in 
% \tablenotetext's.

\clearpage

\clearpage

\figcaption{Average blue spectrum of 0513--69. The resolution is 
$\sim1.3$\,\AA. The principal He{\sc ii} and H emission features are marked, 
along with their associated Doppler-shifted components (see \S~2.2). 
Note the extended blue wing of He{\sc ii}~4686 and lack of blue emission 
in H$\beta$. 
\label{fig1}}

\figcaption{Average red spectrum of 0513--69. The resolution is 
$\sim1.3$\,\AA. The spectrum is dominated by H$\alpha$, but Doppler-shifted 
S$^+$/S$^-$ components are just discernable above the noise. 
\label{fig2}}

\figcaption{The optical light curve of 
0513--69 from 1992 Aug.~22 -- 1995 Nov.~27. These 
observations were obtained as a by-product of the MACHO project (see Alcock 
\etal 1996). The 
relative magnitude is shown for the ``blue'' filter, which is approximately 
equivalent to the Johnson V passband. Note the quasi-regular magnitude drops 
of $\sim 1$~mag. The symbol ``SAAO'' around day no.~1754 indicates the time at 
which we obtained our fast optical photometry (see \S~3.1). 
Downward and upward vertical arrows indicate the times at 
which the system was known to be on (X) and off (NX) in X-rays respectively 
(see \S~1, \S~5.3 and \S~5.4). 
\label{fig3}}

\figcaption{White light photometry of 0513--69 relative to a local standard 
(upper curve). The magnitude of a comparison star relative to the same 
local standard is also plotted (lower curve). 0513--69 shows a significant 
fading of $\sim 0.1$~mags over the duration of the observations (220~mins). 
\label{fig4} }   

\figcaption{Triple Gaussian fit (solid curve) to the He{\sc ii}~4686 
average profile (dash-dotted line). The 
components consist of a narrow function to fit the line core, and 2 broad, 
low-amplitude Gaussians which fit the red and blue sections of the extended 
profile base. \label{fig5} }

\figcaption{CLEANED periodogram for He{\sc ii}~4686 radial velocity 
data (using 10 iterations and a loop gain of 0.1). 
The frequency range is $\nu=0.1-10.0$~cycles~d$^{-1}$, and the search 
increment used is $\Delta\nu=0.001$~cycles~d$^{-1}$. The most power arises 
at P$=0.761\pm0.025$~d, with a one day$^{-1}$ alias at 
$\nu = 0.22$~cycles~d$^{-1}$ 
\label{fig6} }

\figcaption{The CLEANED periodogram for our He{\sc ii}~4686 
velocity data and the radial velocities of C96 
(using 10 iterations and a loop gain of 0.1). The most power arises at 
P$=0.761\pm0.004$~d; one day$^{-1}$ aliases are also present. \label{fig7}}

\figcaption{The radial velocity curve obtained by folding our He{\sc ii}~4686 
data 
on an orbital period of P$=0.76278\pm0.00005$~d, using the photometric 
ephemeris of Alcock \etal (1996). Phase zero corresponds to maximum optical 
light. The parameters of the sinusoidal fit are 
K$ = 14.5\pm3$\,km\,s$^{-1}$, $\gamma = 297\pm1$\,km\,s$^{-1}$ and 
$\phi_{\rm o}=0.17\pm0.09$. Two cycles are plotted for clarity. \label{fig8}}

\figcaption{The component mass parameter space for a mass function of 
0.0002 and inclinations of $3-15^{\circ}$. If the accreting star is a 
white dwarf, the system must be observed nearly pole-on. \label{fig9}}

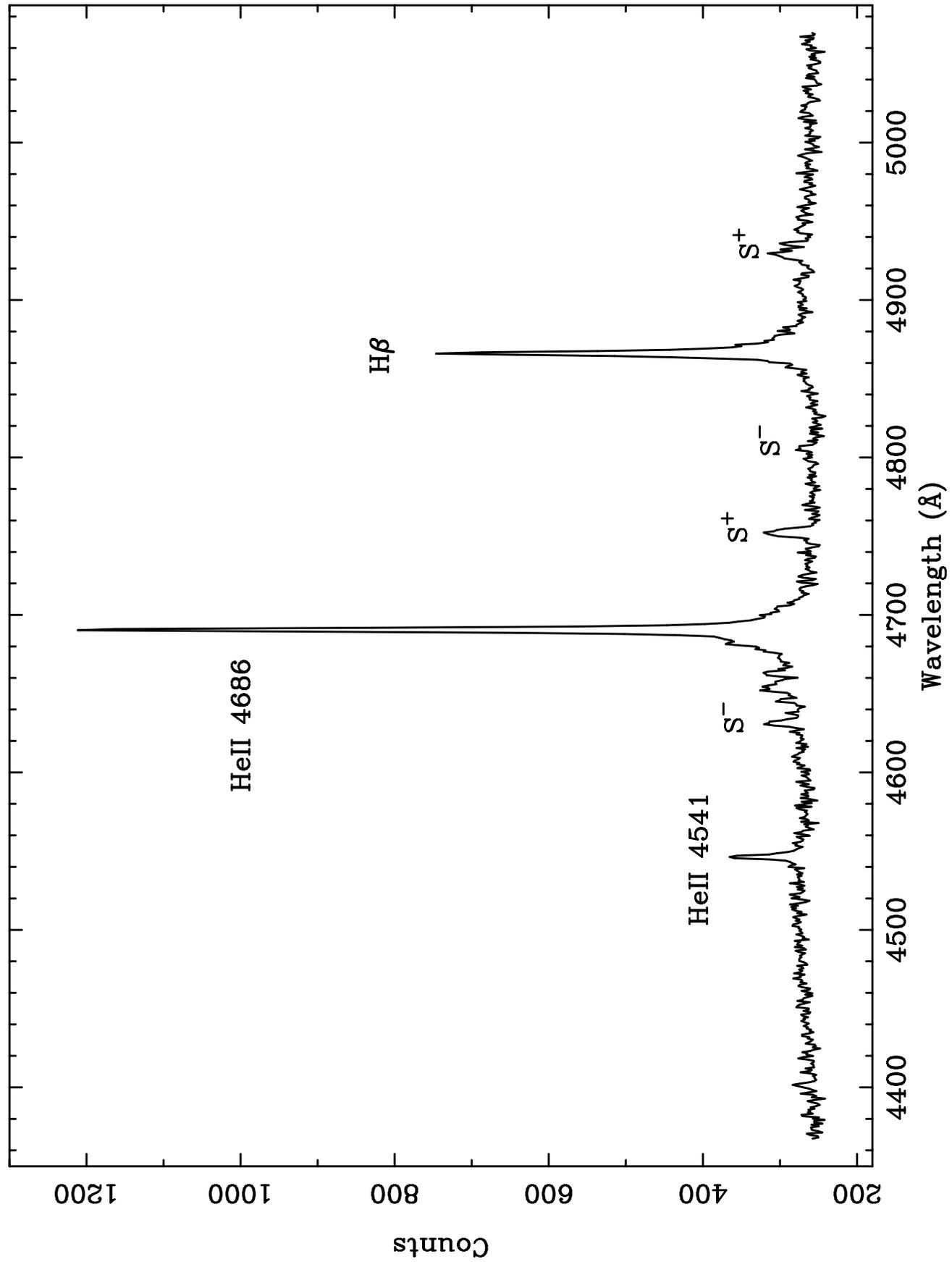
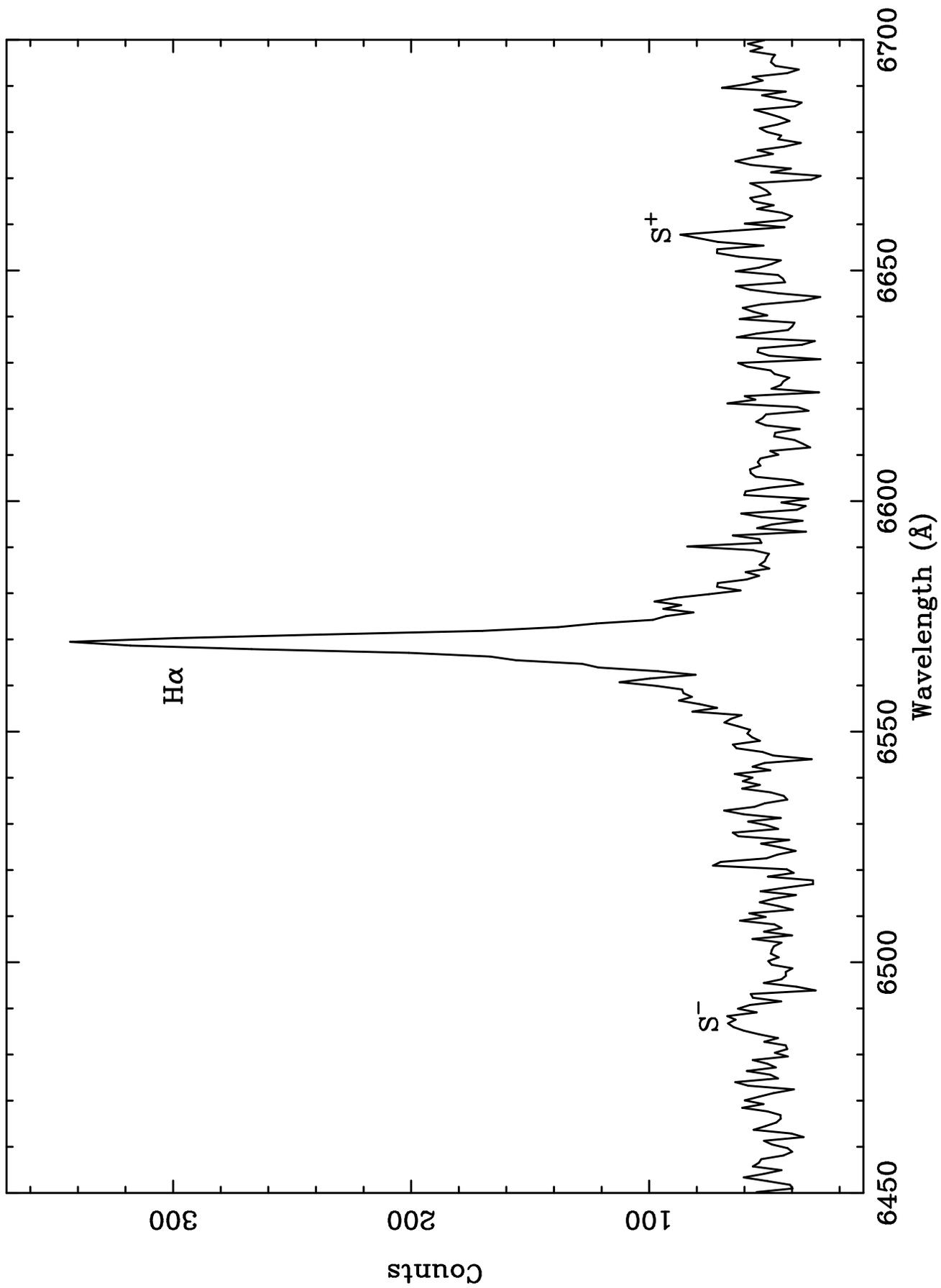
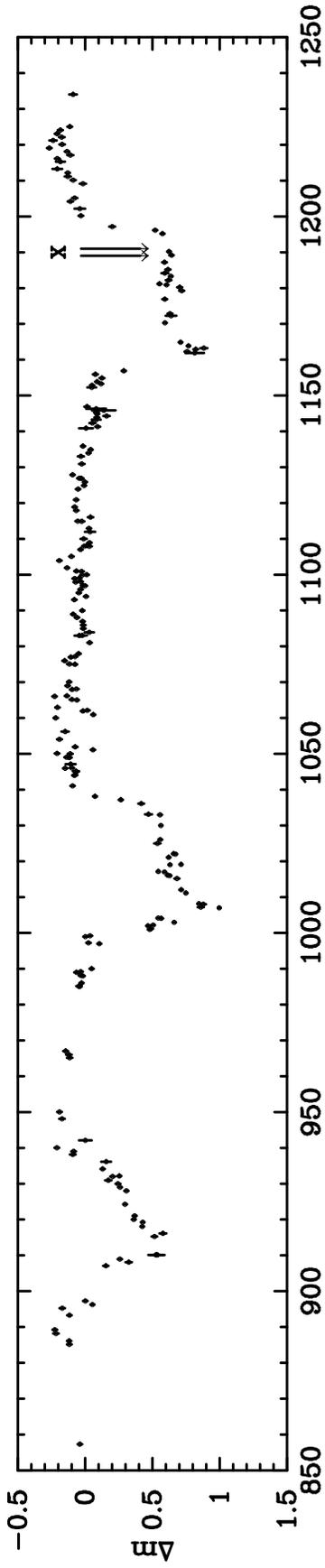
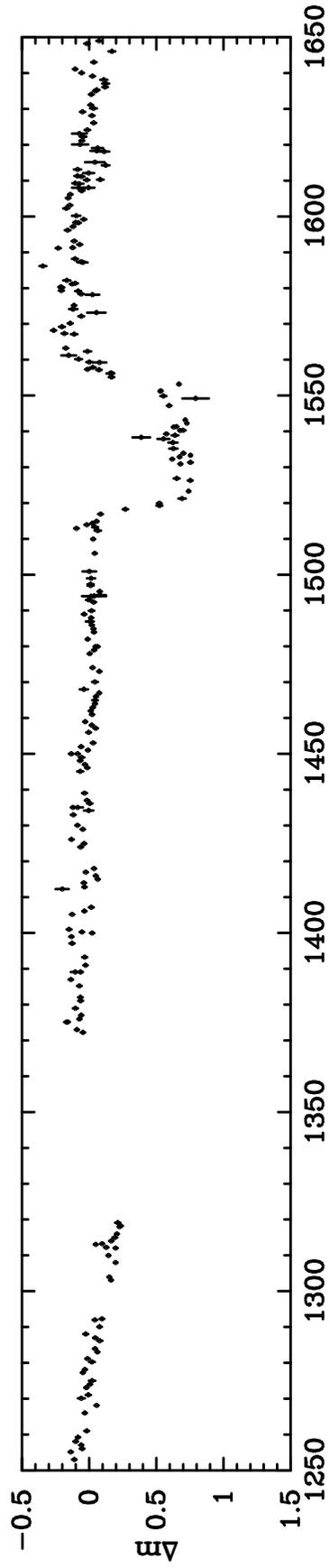
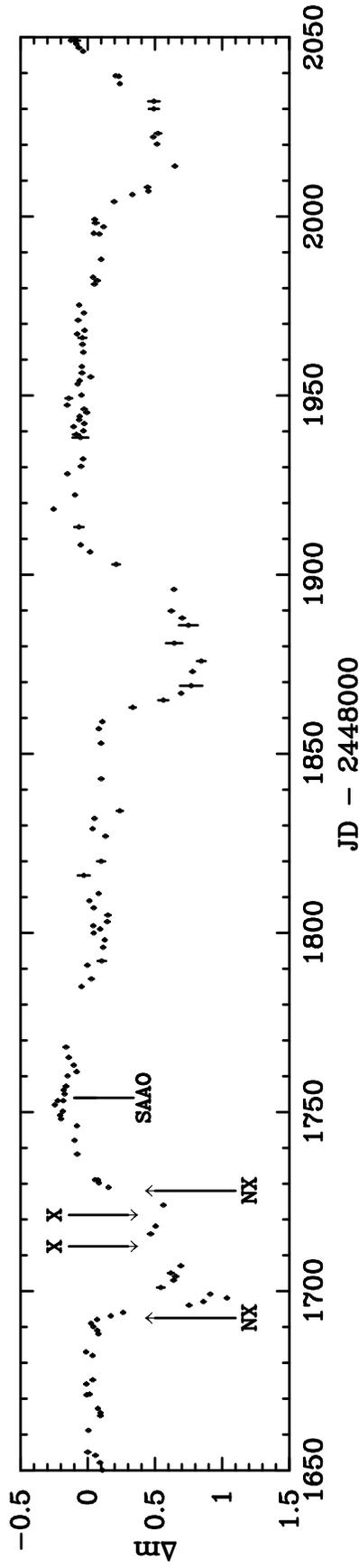
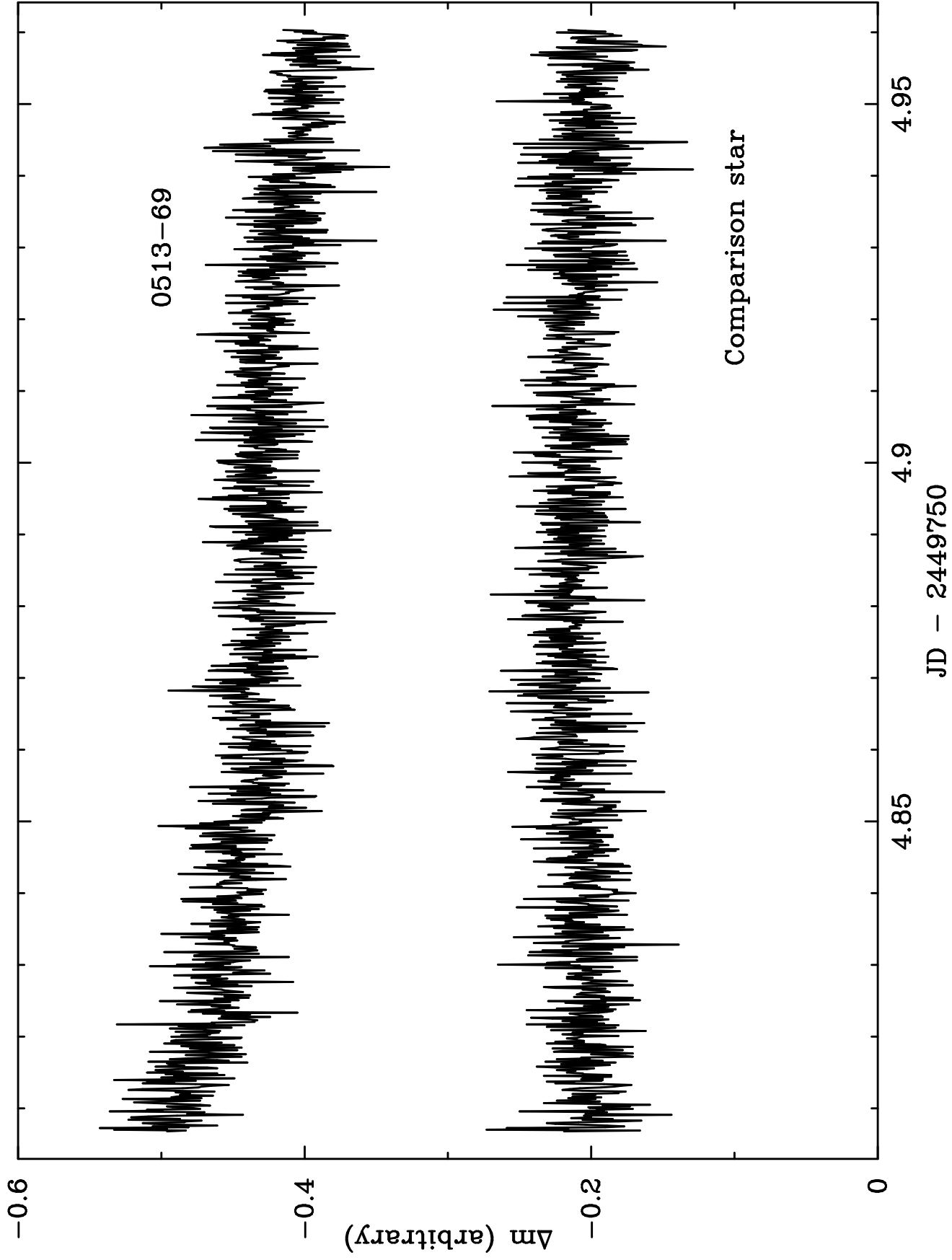
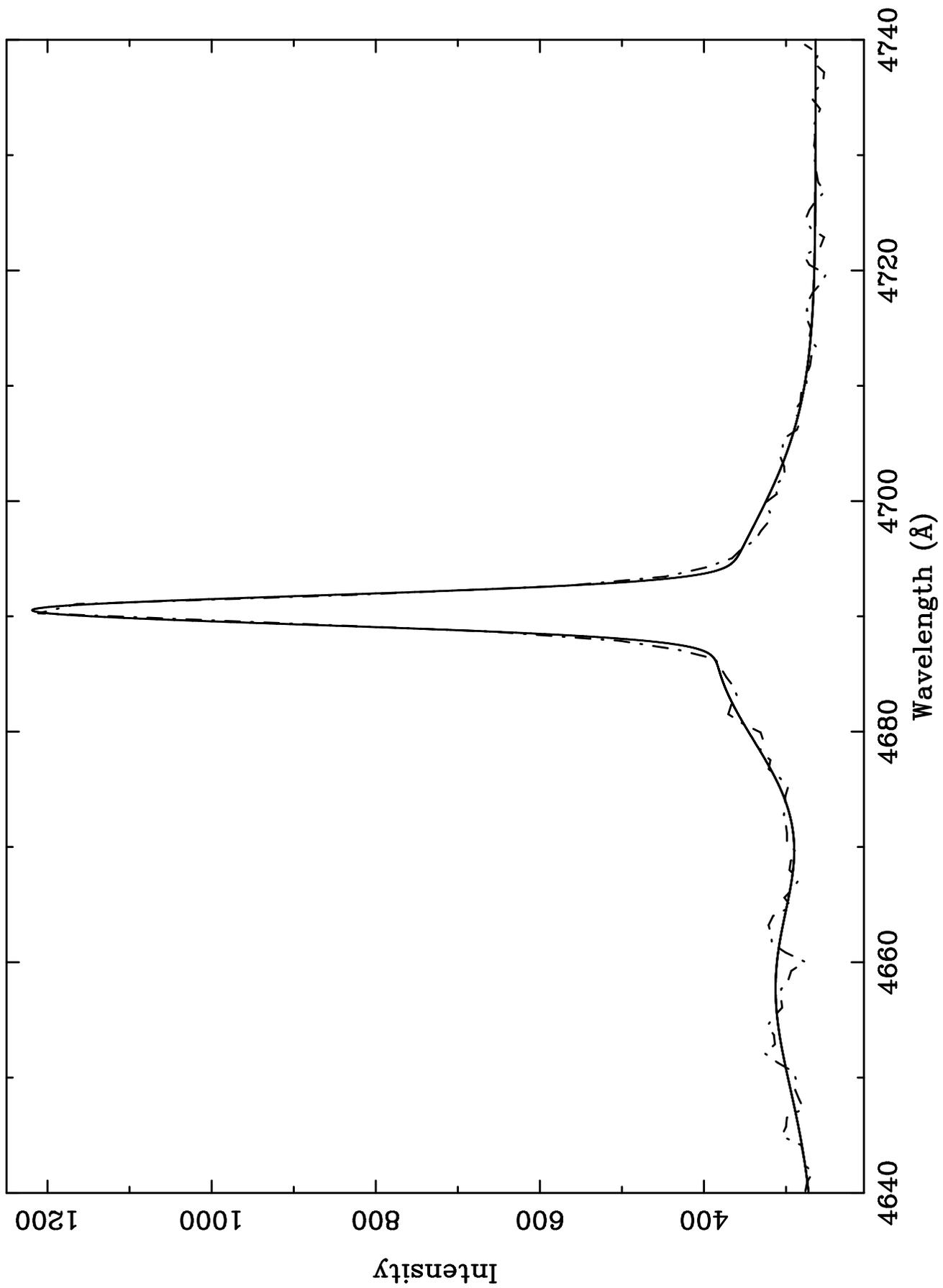
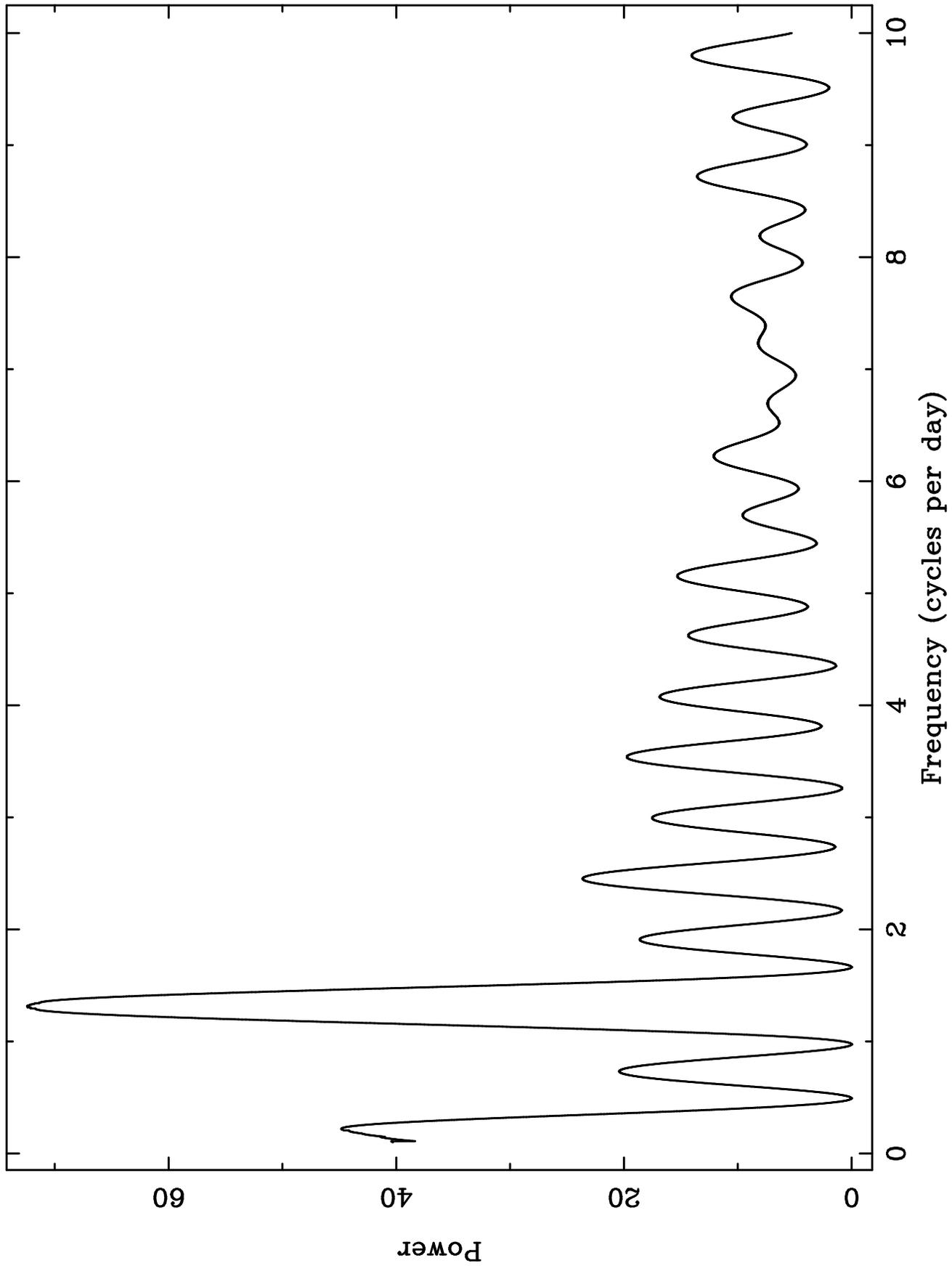
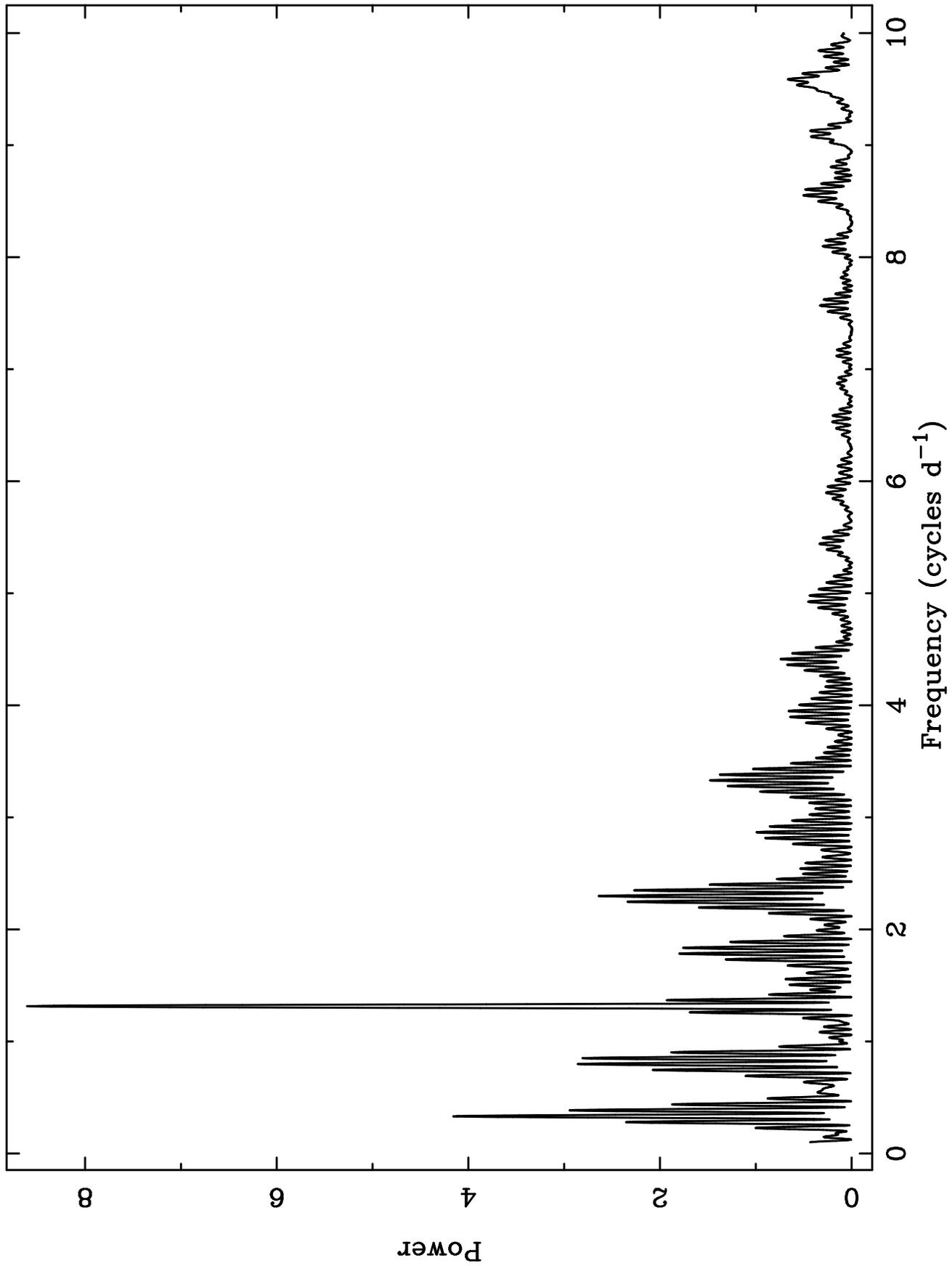
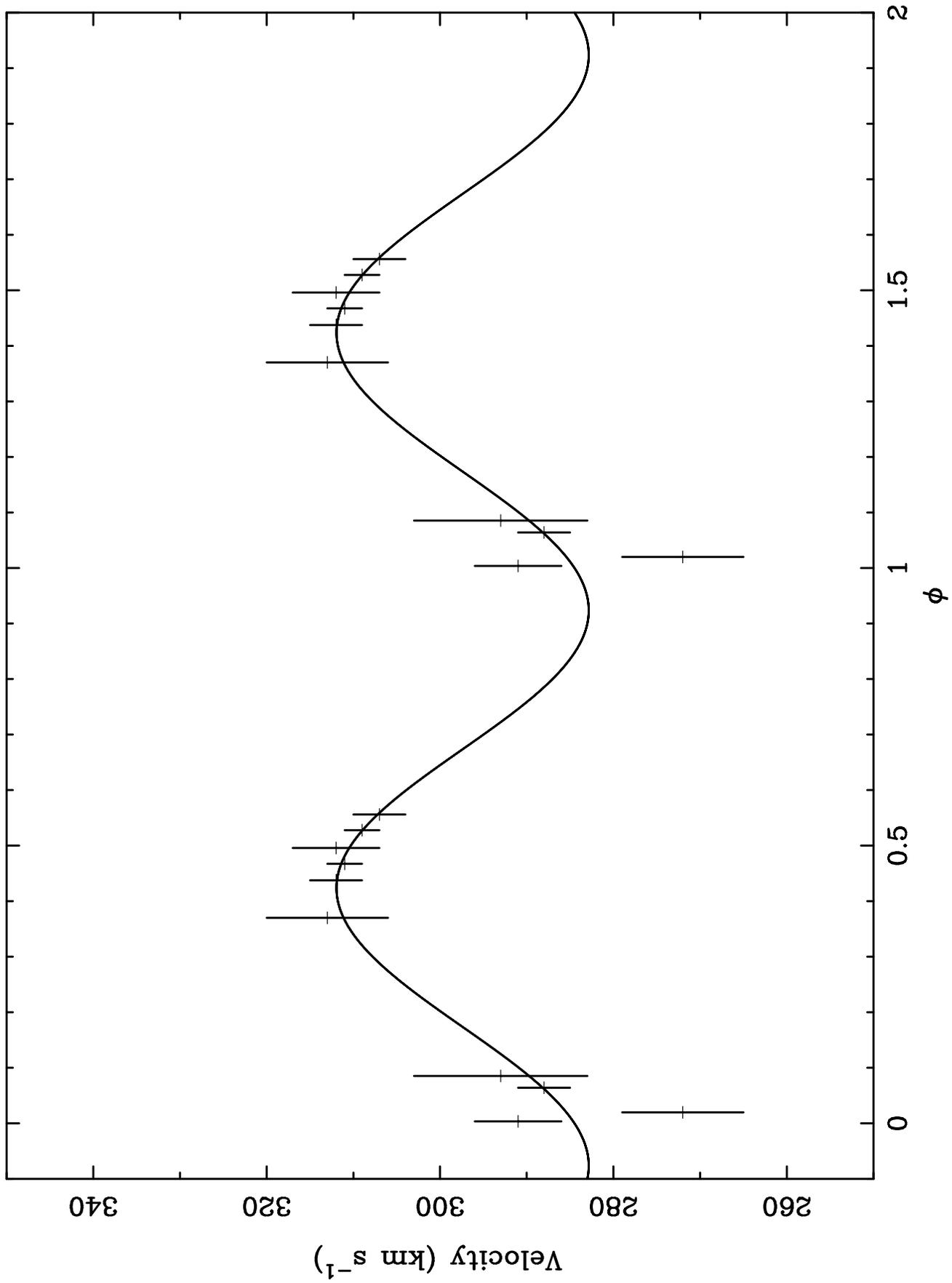
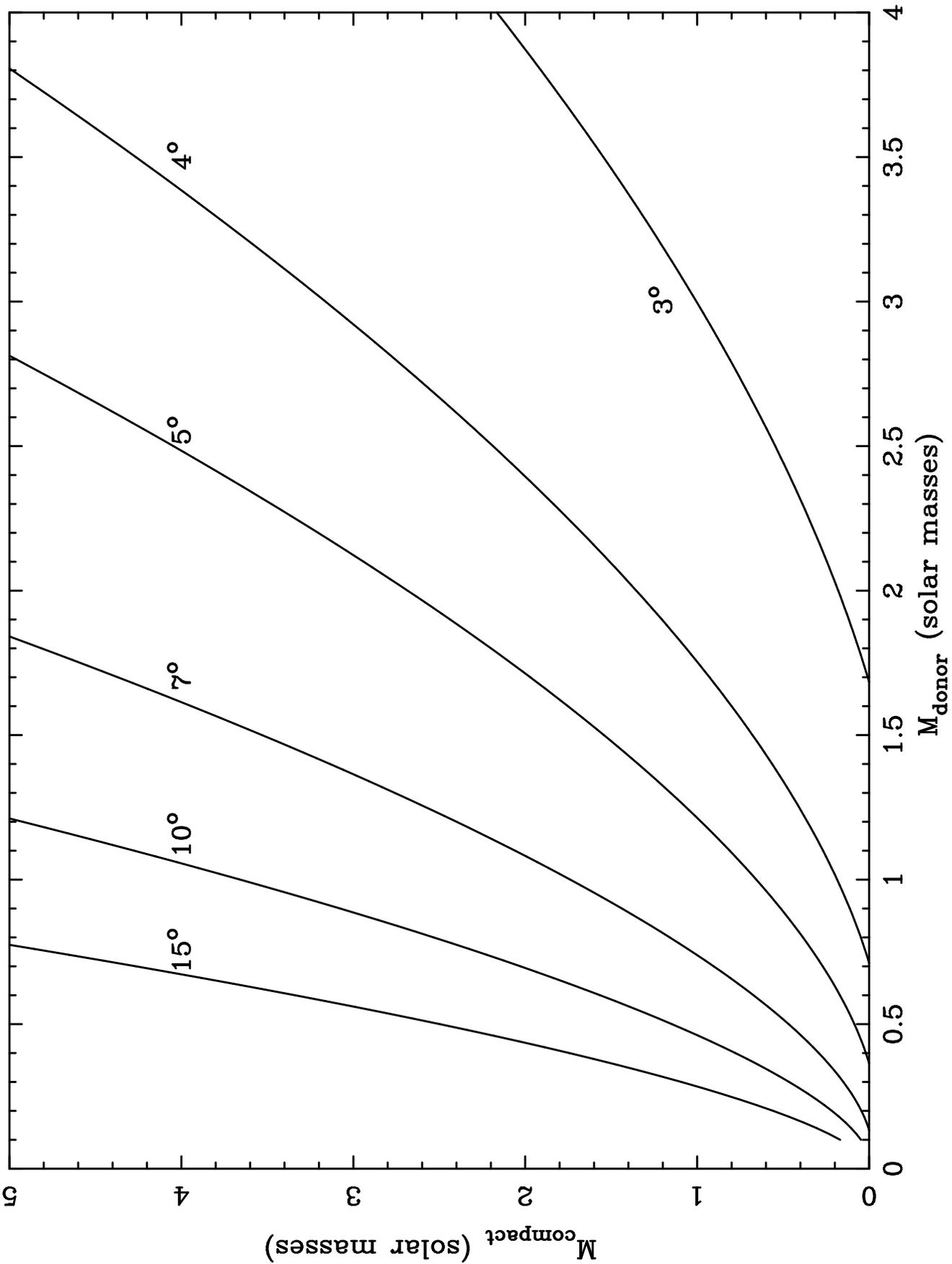
\end{document}